\documentclass[prd,twocolumn,showpacs,preprintnumbers,amsmath,amssymb,nofootinbib]{revtex4}

\usepackage{epsfig}

\def\beq{\begin{equation}}
\def\eeq{\end{equation}}
\def\bea{\begin{eqnarray}}
\def\eea{\end{eqnarray}}
\def\bq{\begin{quote}}
\def\eq{\end{quote}}
\def \lsim{\mathrel{\vcenter
     {\hbox{$<$}\nointerlineskip\hbox{$\sim$}}}}
\def \gsim{\mathrel{\vcenter
     {\hbox{$>$}\nointerlineskip\hbox{$\sim$}}}}
\def\gappeq{\mathrel{\rlap {\raise.5ex\hbox{$>$}}
{\lower.5ex\hbox{$\sim$}}}}
\def\lappeq{\mathrel{\rlap{\raise.5ex\hbox{$<$}}
{\lower.5ex\hbox{$\sim$}}}}

\def\sw2{\sin^2 \theta_W}
\def\s132{\sin^2 \theta_{13}}

\def\Wslash{ \, W  \! \!   \! \! \! / ~ }

\def\dslash{ \, \partial  \! \! \! \! / ~ }

\def\kslash{ \, k   \! \! \! /  }
\def\pslash{ \, p   \! \! \! /   }

\def\m3e{\mu \to e \bar{e} e}
\def\a{\alpha}
\def\b{\beta}
\def\g{\gamma}
\def\d{\delta}

\def\m{\mu}

\def\r{\rho}

\begin{document}


\title{Non-Standard Neutrino  Interactions at Colliders \\ }

\author{ Sacha Davidson}
 \email{s.davidson@ipnl.in2p3.fr}
\affiliation{
IPNL, Universit\'e de Lyon, Universit\'e Lyon 1, CNRS/IN2P3, 
4 rue E. Fermi 69622 Villeurbanne cedex, France}
\author{Veronica Sanz}
 \email{vsanz@yorku.ca}
\affiliation{Department of Physics and Astronomy, York University, Toronto, ON, Canada.}


\begin{abstract}
Many extensions of the Standard Model lead to non-standard neutrino interactions (NSI), 
which can affect the interpretation of current and future neutrino data.
We explore an alternative approach to 
the study 
 of these dimension eight  four fermion interactions
(of coefficient $4 \varepsilon G_F/\sqrt{2}$),
 by bringing into play collider data.
In models where coefficients of potential  dimension six operators
are suppressed by  cancellations,  
LEP2  provides interesting 
bounds on NSI operators 
($\varepsilon \lsim 10^{-2} - 10^{-3}$). 
If NSI are  contact interactions
at LHC energies,
they induce an effective  interaction
  $\overline{q} q W^+W^- e_\a^+ e_\b^-$ 
(according to
 the Equivalence Theorem), 
to which the LHC at 14 TeV and with 100 fb$^{-1}$ of data 
has excellent 
  sensitivity 
($\varepsilon \gsim 3 \times  10^{-3}$). 
\end{abstract}

\pacs{PACS:13.15.+g,14.60.St, 13.85.Rm }
\maketitle

\section{Introduction}
\label{disc}

Many extensions of
the Standard Model, such as Supersymmetry or
leptoquarks, naturally induce  low energy  contact
 interactions of the form:
\beq
\varepsilon_{\a \b}^{fX}  \frac{4 G_F}{\sqrt{2}} 
(\overline{f} \gamma^\r P_X f) (\overline{\nu}_\a \gamma_\r \nu_\b) 
~~~,
\label{eqndisc}
\eeq
where $f \in \{ u, d,e \}$ is a first generation 
charged Standard Model fermion, and
$\a,\b \in \{e, \mu, \tau\}$. These  are referred
to as Non Standard neutrino
Interactions (NSI) \cite{NSI}, and can be
generated by $SU(3) \times SU(2) \times U(1)$ 
gauge invariant operators at dimension
 six or higher.  Future neutrino
facilities, such as a Neutrino Factory \cite{nuFact},
could be sensitive to such
interactions with 
 $\varepsilon \gsim 10^{-4}$\cite{ggggn,Ota:2001pw,KLO}.
Formalism and current bounds \cite{Biggio:2009nt} on NSI are reviewed in
section \ref{notn}.

This paper aims  to make a preliminary  exploration 
of the complementarity 
of  current
collider experiments and future neutrino 
facilities  to 
these neutral current operators. We focus 
on neutral current NSI induced at dimension eight
(because the dimension six operators
are more strictly constrained),
from operators such as  
\beq
\label{eqn2}
\frac{1}{\Lambda_8^4}
(\overline{q} \gamma^\r P_L q) (\overline{H\ell }_\a \gamma_\r H \ell_\b) 
\eeq
where $q$ and $ \ell$ are SM doublets, and  $H$ is the Higgs.
Requiring $\varepsilon \gsim 10^{-4}$ implies
$\Lambda_8 \lsim 2$ TeV. This suggests
that the  new mediating particles
($Z'$s, leptoquarks, sparticles...),
if weakly coupled or contributing
in loops, are kinematically accessible
to the LHC. In this case, their
discovery prospects are 
model-dependent, and have been
widely studied\cite{tdrs}.  Here,
to retain some degree of model independence,
 we  consider effective operators at colliders
(LEP2, the Tevatron, the LHC).  
So for the LHC, 
we are assuming  heavy  New Physics with
$\gsim 1$ couplings.  We also assume 
that the New Physics does not generate
 ``dangerous''
 dimension six operators involving two
charged leptons instead of two neutrinos, because these are
strictly constrained. 
A catalogue of such  models 
 can be found in \cite{slepton,Gavela}.

In section \ref{LEP2}, we argue that in
some models, such as those considered
by Gavela {\it et.al.} \cite{Gavela},
the dimension eight NSI interaction
of the form (\ref{eqndisc}) is accompanied by
dimension eight contact interactions
involving  charged leptons rather than neutrinos,
 with coefficients
$\sim s/\Lambda_8^4, (t-u)/\Lambda_8^4$, where $s$, $t$
and $u$
are the Mandelstam variables.  We estimate
that the LEP2 bounds on such  contact
interactions  could be translated to
$\varepsilon \lsim 10^{-2}, 10^{-3}$. 

In section \ref{LHC}, we use the Equivalence Theorem
to replace $\langle H \rangle \nu_\a \to W^+ e_\a^-$, and study
the prospects for detecting $q \bar{q} \to W^+ W^- \ell_\a^+
\ell_\b^-$ at the LHC. Rough estimates suggest that
couplings $\gsim 1$ are required for 
 NSI  to function as dimension eight contact
interactions at LHC energies (assuming $\varepsilon \gsim 10^{-4}$).
Such contact interactions would induce few events ($\sigma (p p \to
W^=W^- \ell^+\ell^-)  \sim 10^{-3}$  fb $ \times (10^{-4}/\varepsilon)^2$),
 but at very high $p_T$ where Standard Model
backgrounds are negligeable. We estimate that
the LHC could be sensitive to 
$\varepsilon \gsim  3 \times 10^{-3}$.

\section{Review and Notation}
\label{notn}

Non Standard neutrino Interactions (NSI) 
 are  four-fermion interactions
involving at least one neutrino,  which are
induced by Beyond the Standard Model physics. 
For a recent review, see {\it e.g.} \cite{slepton}.
In this paper, we focus on the neutral current NSI,
sometimes called ``matter NSI''.   
At energies $\ll m_W$, the  contact interactions
between a neutrino, anti-neutrino and two first
generation fermions can be written
\bea
{\cal L}_{eff}  &= & {\cal L}_{SM} 
+ {\cal L}_{NSI} \nonumber \\
{\cal L}_{SM}  & = &
   - 2\sqrt{2}G_F
\sum_{P,f,\beta} g^f_{P} [\bar{\nu}_\beta \gamma_\rho P_L
\nu_\beta][\bar{f} \gamma^\rho P f]
\label{SM} \\
{\cal L}_{NSI} & = & - \varepsilon^{fX}_{\alpha \beta} 2 \sqrt{2} G_F
 (\bar{\nu}_\alpha \gamma_{\rho} P_L \nu_\beta) (\bar{f} \gamma^{\rho}P_X
 f)
\label{eps}
\eea
where $f$ is a first generation
fermion, $P_X = P_L$ or $P_R$,
and    the tree-level $g^f_P$ are given in table \ref{tab:gfP}.
We neglect SM loop corrections \cite{sirlin}  in this paper; for a 
discussion which includes them, see \cite{BellPP}.

\begin{table}
$$\begin{array}{|c|cc|}\hline Z\hbox{ couplings} & g^f_L & g^f_R \\
\hline
\nu_e,\nu_\mu,\nu_\tau
& \frac{1}{2} & 0 \\ e,\mu,\tau &-\frac{1}{2}+\sw2 & \sw2 \\ u,c,t &
\frac{1}{2}-\frac{2}{3}\sw2 & -\frac{2}{3}\sw2 \\ d,s,b &
-\frac{1}{2}+\frac{1}{3}\sw2 & \frac{1}{3}\sw2 \\
\hline
\end{array}$$
\caption{$Z$ couplings to SM fermions.
\label{tab:gfP}}
\end{table}

NSI are phenomenologically
 interesting because they  could interfere in  the extraction of neutrino
oscillation parameters  \cite{g,Johnson:1999ci,Ota:2001pw,ggggn},
clouding, for instance the measurement of
$\sin \theta_{13}$
at a $\nu$ Factory
\cite{Huber:2001de}. 
 There have been several studies
of the  sensitivity of current \cite{Ota:2002na} 
and future \cite{future,KLO}  neutrino facilities to 
(charged current) NSI
in the  production and/or detection of neutrinos,
and to neutral current NSI which could  contribute
a ``matter'' effect in neutrino propagation. 
Future facilities, such as a 
$\nu$Factory, could be sensitive
to   $\varepsilon \gsim 10^{-4}$.
Recent bounds on the coefficients of  neutral and
charged current NSI can be found in \cite{Biggio:2009nt},
and are of order  ${\cal O}(1) - 10^{-2}$.

 From a more theoretical perspective, NSI are
also interesting because they would indicate New
Physics in the lepton sector at the  TeV scale.
 However, dimension $>4$ operators induced by New Physics,
such as Eqn.~(\ref{eps}), should be invariant under
the Standard Model gauge symmetries. It is
not simple to construct New Physics
models which would be discovered via NSI.  For instance,
the combination
$(\overline{\nu}_\a \gamma^\rho \nu_\b) $ can easily
be obtained as a component of  
$  (\overline{\ell}_\a  \gamma^\rho  \ell_\b )$. However,
an effective operator containing
this current   also gives rise to contact interactions involving
the charged leptons $(\overline{e}_\a \gamma^\rho e_\b) $,
which are more strictly constrained(see {\it e.g.} \cite{ybook}).  
Neutral current
NSI,  without undesirable charged lepton contact
interactions, can be obtained 
at dimension six from operators such as
\beq
(\overline{\ell} H^*) \dslash (H \ell)
\label{KT}
\eeq
 that modify the
neutrino kinetic terms after spontaneous symmetry
breaking\cite{KT}. 
However,  such modifications to the normalization
of the neutrino kinetic terms  are constrained by the unitarity
of the PMNS matrix, implying 
 that $\varepsilon \lsim 10^{-3}$ \cite{KT}.
Alternatively, a dimension 6 contact interaction
between neutrinos and electrons 
\beq
\ell_{[i} \ell_{j]} (\ell_{[k} \ell_{l]} )^\dagger
\label{slepton}
\eeq 
can be obtained
by 
 the tree-level  exchange of a scalar
with flavour antisymmetric couplings (({\it e.g.} a slepton in
R-parity violating Supersymmetry).
Recent constraints on the coefficient of the operator
(\ref{slepton}) can be found on \cite{slepton}.

As noted in \cite{anna},   gauge invariant operators
 that induce    neutral current NSI,
without accompanying charged lepton contact
interactions, can be obtained at dimension eight,
for instance via the operator
\beq
\frac{1}{\Lambda_8^4}
(\overline{q} \gamma^\r P_L q) (\overline{H\ell }_\a \gamma_r H\ell_\b) 
~~~,
\label{L8}
\eeq
where $q$ and $ \ell$ are SM doublets, and  $H$ is the Higgs.
Requiring $\varepsilon \gsim 10^{-4}$ implies
\beq
\frac{v^2}{\Lambda_8^4}  \gsim \frac{\varepsilon}{v^2}   
\eeq
where $v = \langle H \rangle = 174$ GeV,
so $ \Lambda_8  \lsim 2 $ TeV. The obvious question
arises:  what New Physics could  generate such a large coefficient
at dimension 8 while remaining  consistent
with all other constraints? Various options
have been explored.

\begin{figure}[ht]
\unitlength.5mm
\begin{center}
\includegraphics[scale=.7]{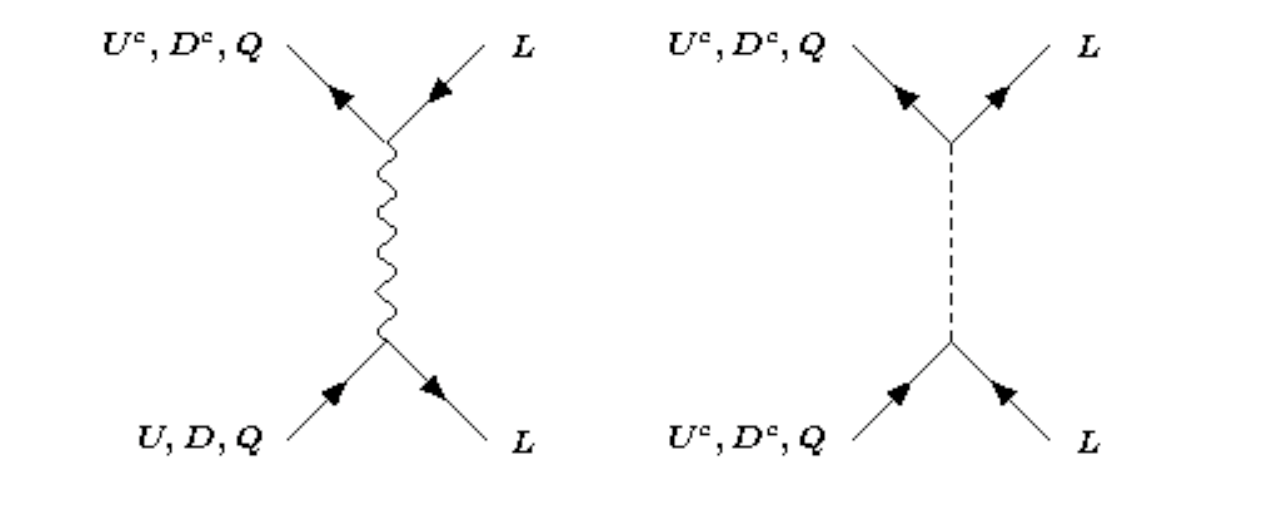}
\end{center}
\caption{ 
Scalar and vector leptoquarks can 
separately   induce 
dimension six   operators involving
two quarks and two charged leptons,
with a relative negative sign.
The mass and  coupling  of the
vector can be tuned relative to the scalar, to obtain
a cancellation at zero momentum transfer,
which allows these leptoquarks 
to participate in generating
significant dimension 8 NSI  operators.}
\label{figdim6} 
\end{figure}

 New Physics 
that induces  the  dimension 8 operators at tree level,
without generating tree level dimension
six operators, was explored in \cite{slepton}.
The models studied in \cite{slepton} also
induced charged current NSI, and non-unitarity,
which restricted the
  $\varepsilon$s  below their phenomenological
bounds. However $\varepsilon \sim 10^{-3} $
could be obtained.    
A less restrictive approach to tree level
New Physics for neutral current NSI was followed
in \cite{Gavela}, where  cancellations
were allowed  in the  coefficient
of the unwanted dimension six operators. For instance,
in a model containing a singlet scalar
and a  singlet vector leptoquark,  of masses
$m_S$ and $m_V$ and couplings to first generation
fermions
\beq
h (\overline{q^c} i \tau_2 \ell)S_0^\dagger 
+ g (\overline{q} \gamma_\mu \ell) V^\mu_2 
\eeq
it is a  straightforward exercise in Fierz
transformations to  show that the sum of
the diagrams in figure  \ref{figdim6} gives
\beq
\left(\frac{g^2}{2m_V^2} -  \frac{h^2}{4m_S^2} \right) {\Big (}
  (\overline{q} \gamma  q)  (\overline{{\ell}} \gamma {\ell}) 
+  (\overline{q} \gamma \vec{\tau}  q)  (\overline{{\ell}} \gamma
\vec{\tau} {\ell}) 
{\Big )} ~~~,
\eeq
so parameters can be chosen such that the coefficient is zero.
This cancellation will no longer be
exact at non-zero momentum transfer.
In section \ref{LEP2} of this paper, we 
show that LEP2 bounds on contact interactions
generically impose $\varepsilon \lsim 10^{-2} - 10^{-3}$ 
on models which circumvent the dimension
6 bounds via such a cancellation. 

New Physics could also induce NSI via
loop diagrams. This option is
disfavoured by  naive power counting:
\beq
\frac{v^2}{16 \pi^2 \Lambda_8^4} \simeq \frac{\varepsilon}{v^2}
\eeq
which implies that such New Physics  should have a 
 mass  $\sim 300$ GeV ---
accessible to current colliders, and 
constrained by precision electroweak data. 
However, the attraction of generating
neutral current NSI at one loop, is
that the SM has a built-in mechanism 
to suppress  
Flavour Changing Neutral Currents
(FCNC) up to dimension
eight:  the (quadratic) GIM
mechanism  (Recall that GIM allows
FCNC with coefficients $\propto m_f^2G_F^2/(16 \pi^2)$,
where $m_f = y_f v$ is the mass of an internal fermion,
so one can interpret that the diagram has
two additional Higgs legs.). 
Minimal Flavour Violation \cite{mfv}
 extends  the GIM mechanism to  New
Physics models, thereby allowing
NP with flavoured couplings  at the TeV
scale, despite the stringent bounds on 
quark FCNC.  So one could anticipate that
NSI will ``naturally'' arise 
via loops in models with 
TeV New Physics in the lepton sector.
An obvious example is Supersymmetry,  which
can have spartner masses in
the $300$ GeV range while respecting
precision elextroweak constraints.
However, supersymmetric
loop contributions to NSI were
computed in \cite{BellPP},
and  it appears difficult to
obtain $\varepsilon \gsim 10^{-4}$.

\section{LEP2 Bounds on Dimension 8 Derivative Operators}
\label{LEP2}

In this section, we estimate bounds from LEP2 on 
contact interactions involving four charged
leptons, with coefficients $\propto \{ s,t,u \}/\Lambda_8^4$.
Some models which give rise to  NSI
also induce such interactions, so
these bounds
may be translated, 
in a  model-
{\it dependent} way,
 to 
the $\varepsilon$ coefficient of 
   NSI involving two neutrinos and
two electrons  ($f = e$ in
eqn \ref{eps}).
In section \ref{LEPops}, we show
that the dimension eight four-charged-lepton
operators can arise
in models where  the  NSI are generated 
at tree level, and  where the
dimension six four-charged-lepton operator
vanishes due to a cancellation (see {\it e.g.}
figure \ref{figdim6}, with the quarks
replaced by electrons).  
Such
models were classified in \cite{Gavela}.
In sections \ref{LEPdiag} and \ref{LEPFV},
we estimate bounds on  $\Lambda_8$.

\subsection{Dimension 8 operators with 4 charged leptons }
\label{LEPops}

Gavela et al \cite{Gavela}  classified  the simplest
extensions of the Standard Model which satisfied
the following conditions:
\begin{enumerate}
\item the  New Physics (NP)  
generates neutral  NSI at tree level
\item
the NP couples to SM bilinears
(this allows to identify the quantum numbers of the 
simplest combination of NP that generates
a dimension  6 or 8 operator)

\item  the   four charged lepton 
operators 
must be suppressed, so 
either the 
NP does  not induce them, or 
the various NP coefficients  sum to zero 
\end{enumerate}

As an example  where the coefficient
of a dangerous four-charged lepton operator
vanishes due to a cancellation, consider a model containing 
an  SU(2) doublet vector of mass $m_V$ and 
an SU(2) doublet scalar  of mass $m_S$ with
couplings
\bea
h  (\overline{e} \ell) \tilde{S}_2^\dagger 
+ 
g (\overline{e^c} \gamma_\mu \ell){V}^\mu_2 
\eea
Neglecting the masses of the external
fermions,  the diagrams of
figure \ref{figdim6} 
give
%
%
the operators
\bea
- \frac{h^2}{m_S^2-t}  (\overline{e} \ell) (\overline{\ell} e) =
 \frac{h^2}{2(m_S^2 -t)}(\overline{\ell} \gamma^\mu \ell) (\overline{e} \gamma_\mu e) 
\nonumber\\
 \frac{g^2}{m_V^2-u}(\overline{e^c} \gamma^\mu \ell) 
(\overline{\ell} \gamma_\mu e^c) 
= -\frac{g^2}{m_V^2-u}(\overline{e} \gamma^\mu e) 
(\overline{\ell} \gamma_\mu \ell) 
\eea
where $t$ and $u$ are the Mandelstam variables.
At zero momentum transfer, the
coefficients of the dimension 6 operator
can be arranged to cancel 
by choosing  $g^2/m_V^2 = h^2/(2m_S^2)$.
However, at dimension 8, 
will arise  operators such as 
\bea
  \frac{g^2(t/m_S^2-u/m_V^2)}{m_V^2}
(\overline{\ell} \gamma^\mu \ell) (\overline{e} \gamma_\mu e) 
\eea
Since $s + t + u = 0$ (in the limit of neglecting
the fermion masses),  
 we take our  double-derivative 
dimension eight operators to be  $\propto s$ and $\propto t-u$:
\begin{widetext}
\bea
\frac{1}{\Lambda_8^4}
( (D^\nu \overline{\ell}_\a) \gamma^\mu (D_\nu \ell)_\b) (\overline{L} \gamma_\mu   L)
  & \to&   \frac{s}{\Lambda_8^4}
(  \overline{e}_\a \gamma^\mu P_L e_\b) (\overline{e} \gamma_\mu P_Y  e)
\label{Lops8}
\\
\frac{1}{\Lambda_8^4}
(( D^\nu \overline{\ell}_\a) \gamma^\mu \ell_\b) (\overline{L} \gamma_\mu  (D_\nu L))
-
\frac{1}{\Lambda_8^4}
(\overline{\ell}_\a \gamma^\mu (D^\nu \ell)_\b) (\overline{L} \gamma_\mu  (D_\nu L)) 
 & \to &  \frac{t-u}{\Lambda_8^4}
(  \overline{e}_\a \gamma^\mu P_L  e_\b) (\overline{e} \gamma_\mu P_Y  e)
\label{Lopt8}
\eea
\end{widetext}
where $L$ is a first generation lepton,   either a doublet ($P_Y = P_L)$
   or singlet($P_Y=P_R$),
and to the right of the arrow is the four charged lepton
component of the operator on the left.

\subsection{Bounds on flavour diagonal lepton operators}
\label{LEPdiag}

 LEP2 searched for dimension six
 contact interactions  of the form
\beq
\pm \frac{4 \pi}{\Lambda_{6,\pm}^2} 
(\overline{e} \gamma^\mu P_X e) (\overline{f}_\a \gamma_\mu  P_Y f_\a) 
\label{d6}
\eeq
in the channels
$e^+e^- \to e^+ e^-,  \mu^+ \mu^-,  \tau^+ \tau^-,  \bar{q} q$
(where $q \in \{u,d,s,c,b \}$ is a light quark; we postpone
the quark case to section \ref{qqll}), 
by comparing  the observed  cross-section 
\beq
\sigma = \int_{1}^{-1}  d c_\theta \frac{d \sigma}{dc_\theta} 
\eeq
and forward-backward asymmetry
\beq
A_{FB} = \frac{ \int_{0}^{-1}  dc_\theta \frac{d \sigma}{dc_\theta} 
-\int_{+1}^{0}  dc_\theta \frac{d \sigma}{dc_\theta} }{ \sigma}
\label{defnsAFB}
 \eeq
 to
SM expectations. 
 $c_\theta = \cos \theta$ is the angle between the incoming
and outgoing fermions. 

To translate the published bounds
on $\Lambda_{6,\pm}$
 into constraints
on  dimension eight operators $\propto s/\Lambda_8^4$,
we focus on LEP2 data at large $\sqrt{s} :183 \to 208$ GeV. 
As there was  significant initial state radiation, 
several experiments  required that the invariant
mass of the final state leptons exceed 85\% of
the centre of mass beam energy. So we approximate
\beq
\frac{4 \pi}
{\Lambda_6^2} \simeq   \frac{ s}{  \Lambda_8^4}
 \simeq   \frac{ v^2}{  \Lambda_8^4}
\label{quesuisje1}
\eeq
where $v = \langle H \rangle = 174$ GeV. If we assume
that the  NSI operator with
coefficient $v^2/\Lambda_8^4$ is accompanied
by the four charged lepton operator with
coefficient  $s/\Lambda_8^4$,
then  we can extrapolate bounds  of order
\beq
\varepsilon \lsim \frac{4 \pi v^2 }
{\Lambda_6^2} \simeq 3.8 \times 10^{-3}\frac{(10 TeV)^2 }
{\Lambda_6^2}
\label{guess}
\eeq

Consider now the dimension eight four charged
lepton operator with coefficient $(t-u)/\Lambda_8^4$,
eqn (\ref{Lopt8}).
Since
\beq
t-u = c_\theta s
\eeq 
it is  clear from eqn (\ref{defnsAFB}) that
this operator will have  different contributions
 to $\sigma$ and $A_{FB}$ than  those
of eqns (\ref{Lops8}) and (\ref{d6}),
because  even powers of
$c_\theta^2$ in  the matrix-element-squared  contribute to $\sigma$,
and odd powers to $A_{FB}$.

We focus  on  the
interference term
 ${\cal M}_{SM} {\cal M}_{\Lambda}^* + h.c.$
between  the SM and the  contact interaction\footnote{This term is generically larger than  $|{\cal M}_{\Lambda}|^2$, for $\Lambda_6$ of order the experimental bounds.},
and  we treat it as  a small
correction to the SM contribution.
If  we define the total cross-section
in the presence of an operator of the
type  (\ref{Lops8}) to be $\sigma =
\sigma_{SM} (1 + \Delta \sigma_s)$, and
$A_{FB} = A_{FB,SM} (1 + \Delta A_{FB,s})$,  then
some straightforward algebra\footnote{Recall  that, for massless external fermions, $t = -s(1 - c_\theta)/2$ and $u = -s(1 + c_\theta)/2$.   To obtain a term in $|{\cal M}|^2$ which contributes to $A_{FB}$
( is linear in
$c_\theta$), from an
operator of the form eqn (\ref{Lops8}) or
(\ref{d6}), requires  the contraction of a 
$\varepsilon_{\a \b\g\d}$ from  both the Dirac
Traces on the initial and final state fermions. 
Since  $\varepsilon_{\a \b\epsilon \d}$ arise from
Tr$\{ \g_\a \g_\b \g_\epsilon \g_\d \gamma_5 \}$,
it is clear that  $\sigma \propto (g^{e 2}_V + g^{e 2}_A)
(g^{\ell 2}_V + g^{\ell 2}_A)$
and $A_{FB} \propto g^e_V g^e_A
g^\ell_V g^\ell_A$, where $\ell$ is the final
state lepton. For operators $\propto (t-u)$ of
the form of eqn (\ref{Lopt8}), these relations
are reversed:  $A_{FB} \propto
 (g^{e 2}_V + g^{e 2}_A)
(g^{\ell 2}_V + g^{\ell 2}_A)$
and $\sigma \propto g^e_V g^e_A
g^\ell_V g^\ell_A$. Expanding the various
$g_{V,A}^i$ couplings as a SM part plus a NP
part gives equations (\ref{simple}).} shows that
\bea
\Delta \sigma_{t-u}& = &\pm \frac{\Delta \sigma_{s}}{4} \\
\Delta A_{FB, t-u}& = &\pm 4 \Delta A_{FB,s} 
\label{simple}
\eea
where $+$ and $-$  are respectively for  for $P_Y = P_L, P_R$
in eqn  (\ref{Lopt8}).
So to  obtain bounds on  four-charged-lepton-operators 
with coefficients
$\propto  (t-u)/\Lambda_8^4$, we should refit
to the data. Instead, since the LEP2
data agrees with the SM, we assume that
the bounds on the  $(t-u)/\Lambda_8^4$ operators 
of eqn   (\ref{Lopt8}) are comparable to the
bounds on  the $s/\Lambda_8^4$ operators 
of eqn   (\ref{Lops8}).

Using   bounds  on $\Lambda_{6}$  obtained by
  ALEPH \cite{ALEPH} 
and  OPAL \cite{OPAL},
and by  Bourilkov  
from a  combination of  data from the four experiments
\cite{bourilkov},   we use  eqn
(\ref{guess})
to  estimate the bounds in table
\ref{tab:LEP2bds}.

\begin{table}
\begin{tabular}{||c|c|c||}
\hline
$(\overline{e} \gamma^\mu P_X e) (\overline{\ell} \gamma_\mu  P_Y \ell)
$&  bound & $\varepsilon$ \\
\hline
$e^+ e^- \to e^+ e^-$ &&\\
XY=LL& $\Lambda_{6+} \gsim 10.3$ TeV
& $\lsim 3.7 \times 10^{-3}$ \\ 
~~~LL&$ \Lambda_{6-} \gsim 8.3$ TeV
& $\lsim 5.6 \times 10^{-3}$ \\ 
~~~RL & $\Lambda_{6+} \gsim 8.8$ TeV
& $\lsim  4.7  \times 10^{-3}$ \\ 
~~~RL & $\Lambda_{6-} \gsim 12.7$ TeV
& $\lsim 2.4 \times 10^{-3}$ \\ 
\hline
$e^+ e^- \to \mu^+ \mu^-  $&&\\
XY=LL & $\Lambda_{6+} \gsim 8.1$ TeV 
& $\lsim 5.9 \times 10^{-3}$ \\ 
~~~LL &$ \Lambda_{6-} \gsim 9.5$ TeV  
& $\lsim 4.3 \times 10^{-3} $\\ 
~~~RL &$ \Lambda_{6\pm} \gsim 6.3$ TeV  
& $\lsim 9.1 \times 10^{-3}$ \\ 
\hline
$e^+ e^- \to \tau^+ \tau^-$  &&\\
XY=LL & $\Lambda_{6+} \gsim 7.9$ TeV  
& $\lsim 6.2 \times 10^{-3}$ \\ 
~~~LL & $\Lambda_{6-} \gsim 5.8$ TeV  
& $\lsim 1.1 \times 10^{-2}$ \\ 
~~~RL & $\Lambda_{6+} \gsim 6.4$ TeV  
& $\lsim 9.1 \times 10^{-3}$ \\ 
~~~RL & $\Lambda_{6-} \gsim 4.6$ TeV  
&$ \lsim 1.8 \times 10^{-2}$ \\ 
\hline
\end{tabular}
\caption{The middle column gives the  LEP2 bounds \cite{bourilkov,OPAL,ALEPH} 
on the dimension six contact interaction of the first column, normalized
according to eqn (\ref{d6}).  These bounds can be
used to constrain dimension  eight 
operators with two derivatives and
four charged leptons. If we assume that NP inducing
the derivative operators also induces NSI, we obtain  
a bound  on the $\varepsilon$ coefficient of  the 
dimension eight NSI  operator, which is 
obtained from eqn (\ref{quesuisje1}), and given
in the third column. Although the bounds
are quoted with two significant figures, they
are merely order of magnitude estimates,
as various constants could appear in the passage
between four-charged-lepton-derivative  operators 
and  NSI operators.
\label{tab:LEP2bds}}
\end{table}

\subsection{ Flavour off-diagonal lepton operators}
\label{LEPFV}

The OPAL experiment saw one $ e^+e^- \to e^\pm \mu^\mp$
 event at $\sqrt{s} = 189-209$ GeV, and published limits
\cite{OPALFV}
on $\sigma( e^+e^- \to e^\pm \mu^\mp, e^\pm \tau^\mp,  \tau^\pm \mu^\mp)$
at LEP2 energies. These bounds give more
stringent limits  on operators of the form
(\ref{Lops8}) and (\ref{Lopt8})  than 
the LEP1 bounds on lepton flavour-changing $Z$ decays. 
For  $\sqrt{s} = 200-209$ GeV, OPAL
found
\beq
\sigma  (e^+e^- \to e_\a^\pm e_\b^\mp)
< \left\{ \begin{array}{cc}
22~ fb  &[e \mu]  \\
78~ fb  &[e \tau]  \\
64~ fb  &[\tau \mu]  
\end{array}
\right.
\eeq 
We estimate that the cross-section induced
by the operators of eqns  (\ref{Lops8}) 
and  (\ref{Lopt8})  is  
\beq
\sigma_8 \simeq \frac{s^3}{32 \pi \Lambda_8^8} \times \left\{ \begin{array}{cc}
1 &   , t-u\\
8/3& , s
\label{estdim8}
\end{array}
\right.
\eeq 
and obtain the  bounds on
$\Lambda_8$ listed in  table \ref{tabFV}
(we neglect the 8/3,  so estimate the bound from
$\sigma_8 \simeq s^3/(32 \pi \Lambda_8^8)$.).
 If we assume that
the
 four-charged-lepton-operators with
coefficients $\propto s/\Lambda^4_8, (t-u)/\Lambda^4_8$,  
would be accompanied by
an  NSI operator with coefficient
$v^2/\Lambda^4_8$, 
then we  can  translate
to bounds  on $\varepsilon = v^4/\Lambda_8^4$,
which are given in the last column  of table 
 \ref{tabFV}.

\begin{table}
\begin{tabular}{||c|c|c||}
\hline
$(\overline{e} \gamma^\mu P_X e) (\overline{\ell} \gamma_\mu  P_Y \ell)
$&  bound & $\varepsilon$ \\
\hline
$e^+ e^- \to e^\pm \mu^\mp$ &&\\
$\forall~ XY$ & $\Lambda_8 \gsim 570$ GeV
& $\lsim 8.7 \times 10^{-3}$ \\ 
\hline
$e^+ e^- \to e^\pm \tau^\mp$ &&\\
$\forall~ XY$ & $\Lambda_8 \gsim 485 $ GeV
& $\lsim 1.6 \times 10^{-2}$ \\ 
\hline
$e^+ e^- \to \tau^\pm \mu^\mp$ &&\\
$\forall~ XY$ & $\Lambda_8 \gsim 500 $ GeV
& $\lsim 1.5 \times 10^{-2}$  \\
\hline 
\end{tabular}
\caption{Bounds on  the scale of
dimension eight four-charged-lepton
derivative operators, normalized as
eqns (\ref{Lops8}) and (\ref{Lopt8}), from OPAL
constraints\cite{OPALFV}  on lepton flavour-changing contact interactions
($e^+e^- \to e^+_\alpha e^-_\beta$, $\a \neq \b$). 
These bounds come from  $|{\cal M}_{\Lambda} |^2$,
so are the same for all fermion chiralities and
are independent of the sign of the contact interaction
coefficient. 
 The bounds  on
$\varepsilon$  given in
the last colomn are  estimated by 
assuming that  the dimension eight NSI operator
is induced with  an ${\cal O}(1)$ coefficient at
the same scale.
\label{tabFV}}
\end{table}

\subsection{ Two-quark, two lepton operators}
\label{qqll}

LEP2 set bounds
on dimension six contact interaction of the form 
of eqn(\ref{d6}), involving two electrons and
two light quarks $q \in \{ u,d,s,c,b \}$. 
The bounds  \cite{ALEPH}  were obtained 
assuming the same contact interaction  to all
quark flavours but the top. Nonetheless,  we 
estimate that we can apply them
separately to contact interactions
involving $u$ or $d$ quarks. This is
because  the  bounds arise 
mostly from the interference of
the dimension six contact interaction  
with the SM matrix element, and 
the  cancellations in this flavour sum 
make it of the same order for one or five flavours.
The ALEPH \cite{ALEPH} bounds  on dimension
six contact interactions 
are listed  in  
the second colomn of table \ref{tab:qqll}.
We  can translate these
into bounds on  the  dimension
eight double derivative operators
of eqns (\ref{Lops8},\ref{Lopt8})
by approximating  $s \sim v^2 \sim t-u$, 
as discussed in section \ref{LEPdiag}
for lepton final states. 
  We guestimate
the bounds  using 
the average of $\Lambda_{6+}, \Lambda_{6-}$,
because the effect of the sign would be different
for $u$ and $d$ type quarks. 
If we further assume that the dimension eight
double-derivative operators are accompagnied
by the NSI operator with a coefficient of
the same order, then we can 
estimate bounds on $\varepsilon$
using eqn (\ref{quesuisje1}), which are
given in the last colomn of 
table \ref{tab:qqll}.

\begin{table}
$~$\\
\vspace{.3cm}
\begin{tabular}{||c|c|c||}
\hline
$(\overline{e} \gamma^\mu P_L e) (\overline{q} \gamma_\mu  P_Y q)
$&  ALEPH bound on $\Lambda_6$ & $\varepsilon \lsim $ \\
\hline
Y=L 
&$ \Lambda_6 \gsim 8.0_+, 9.7_-~ $TeV &$ \lsim 4.8  \times 10^{-3} $\\ 
 Y=R &$ \Lambda_6 \gsim 5.2_+,4.1_-~$ TeV  & $\lsim  1.8 \times 10^{-2}$ \\ 
\hline
\end{tabular}
\caption{The second colomn gives 
bounds from ALEPH \cite{ALEPH}
 on  the scale of dimension six 
contact interactions operators, normalised as
eqn  (\ref{d6}). 
The estimated  bounds on $\varepsilon$
given in the last colomn are obtained using
eqns (\ref{quesuisje1}) and  (\ref{guess}).
\label{tab:qqll}}
\vspace{-.4cm}
\end{table}

 Bounds on contact interaction  involving two
muons and two first generation quarks,
normalised according to  eqn (\ref{d6}),
were obtained 
at the Tevatron by CDF \cite{Abe:1997gt},
who searched for excesses in the lepton 
mass spectrum in  110 $pb^{-1}$ of data. 
Using 1 $fb^{-1}$  of data, 
D0 \cite{D0modindep}  performed a ``model-independent''  search for
new physics in a wide range of final states, including
$e^\pm \tau^{\mp}$,$e^\pm \mu^{\mp}$
and $\mu^\pm \tau^{\mp}$, and did not
find evidence for contact interactions. 
It is difficult to extrapolate either
the CDF or D0 bounds to  dimension eight
double-derivative contact interactions. 
CDF obtained bounds $\Lambda_6 \gsim 3-4$ TeV
for $(\bar{u} \gamma^\sigma P_X u)$$(\bar{\mu} \gamma_\sigma P_Y \mu)$ 
interactions, and 
 $\Lambda_6 \gsim 1.5- 2.3$ TeV
for $(\bar{d} \gamma^\sigma P_X d)$$(\bar{\mu} \gamma_\sigma P_Y \mu)$ 
 interactions. We calculated
\footnote{with CTEQ10 parton distribution
functions}
 the contribution to  
  $\sigma(\bar{p} p \to \bar{\mu } \mu) $ arising in the 
presence of  a contact interaction with coefficient $\hat{s}/\Lambda_8^4$,
and compared it to the contribution  due
to a contact interaction with a coefficient  
$4 \pi/\Lambda_6^2$. Requiring that  the dimension
eight contribution be less than  the allowed
dimension six contribution gives  $\Lambda_8 \gsim 500(200)$ GeV,
for  contact interactions involving  $u$($d$) quarks. 
If one makes the further assumption
that the dimension eight double-derivative-operators
accompany  NSI operators, this suggests that
 bounds  of order $\varepsilon \lsim 1 -.01$
could be obtained for NSI involving first generation
quarks and any leptons other than $\bar{\tau} \tau$. 

\section{ LHC discovery reach}
\label{LHC}

In this section, we study the prospects
at the LHC for  NSI involving quarks, in the improbable,
but compartively model-independent,
scenario that they appear as contact
interactions.   We first argue  in section
\ref{eqthm} that 
NSI can give a final state with $W^\pm$
and charged leptons, then we estimate
the production cross-section at the 14 TeV LHC, and
finally discuss sensitivity and backgrounds in
section \ref{eff}.

One might   ``expect''  the  new particles 
which mediate NSI to  be kinematically
accessible to the LHC, so they would not
induce contact interactions.
Consider first   NSI  which arise  at one  loop, 
then $\varepsilon \sim v^4/(16 \pi^2 \Lambda^4)
\gsim 10^{-4}$  implies that 
the new particles  of mass $< \Lambda$
are light enough to be produced at the LHC.
Alternatively, if NSI arise  at tree level,
 then  $\varepsilon \sim v^4/ \Lambda^4
\gsim 10^{-4}$  implies  $\Lambda \lsim 2$ TeV. 
The new particles mediating NSI on quarks
must couple to first generation quarks,
so they must have a large mass, beyond the
LHC reach, to avoid being directly
produced.
The LHC reach for mediators
such as $Z'$s or  leptoquarks, depends
on the details of  the particular model,
but  is of order 3-5 TeV \cite{tdrs}.
However, more
realistically,  the parameter $1/\Lambda^4$
is   a product  of coupling
constants divided by a product of masses,  so
masses beyond the LHC reach could arise
for   new couplings  $>1$,
or if only  some of the new particles  are
``sufficiently'' heavy  ( {\it e.g.} 
$\Lambda^4 = M^2 m^2$, $M \gg m$).
In section \ref{secZ'}, we discuss
 a model with couplings $\leq 1$ and a hierarchy
in masses, and show
that the cross-section can be enhanced,
because    some of the new particles
can be produced on-shell so
 there is less  phase space suppression.
Despite the first sentence
of this paragraph, we assume NSI parameters such  that 
the only signal at the LHC would
be a contact interaction.

\subsection{The Equivalence Theorem and NSI}
\label{eqthm}

In a gauge invariant dimension eight NSI
operator,
the  $H_0^* H_0 \bar{\nu}_\a \nu_\b$
interaction is accompagnied by
$H^+ H^- \bar{e}_\a e_\b$, which
could be expected to reincarnate, after electroweak symmetry
breaking, as
a vertex involving  $W^+ W^- \bar{e}_\a e_\b$.   
This expectation
can be formalised, at energies $\gg m_W$,
via the Equivalence Theorem\cite{EqT,which}, which 
identifies the Goldstone $H^\pm$
with  the longitudinal  component
of the $W^\pm$.

Intuitively, the Equivalence Theorem
relies on the similarity, 
 at energies $\gg m_W$, of  the timelike
$\propto  (E,0,0, k)/m_W$ 
and longitudinal $\propto  (k,0,0, E)/m_W$ 
 polarisation states of the $W^\pm$ gauge bosons.
Since  the  time-like polarisation is
``unphysical'',  and  cancelled by
the goldstone, this  suggests that  the
longitudinal $W^+_L$ can be replaced by the goldstone.

More formally,   after fixing the gauge of the
 electroweak  sector, 
there are remaining
global symmetries, which give
BRS conditions, such as (in Feynman t'Hooft gauge) :
\beq
\partial_\mu W^{+\mu} -  m_W H^+ = 0 ~~.
\label{eqt}
\eeq 
These relations  generate
the electroweak  Slavnov-Taylor Identities. 
We interpret the Equivalence  Theorem 
to say \cite{which}  that this condition can be  imposed
 on physical matrix elements. 
In particular, in an NSI operator,
 we would like to replace
$\nu_\a \langle H \rangle \to  W_L e_\a$.

As an example of how the Equivalence Theorem
works, it is instructive to  consider a one
generation type I seesaw model, containing
a heavy singlet majorana neutrino $N_R$,
of mass $M_R \gg v$,  with
a Yukawa interaction $ \lambda \overline{N_R} H \ell$,
where $\lambda = 1$. If SU(2) is unbroken,
it is straightforward to verify \cite{PRep} 
that $N_R$  has equal
branching ratios to $H^0 \nu, H^{0*} \overline{\nu},
H^+ e^-$ and $ H^{-} e^+$.  After electroweak
symmetry breaking,
the heavy mass eigenstate becomes
\beq
\label{Leig}
L \simeq N_R + \frac{\lambda v}{M_R} \nu_L^c ~~.
\eeq
Despite  being mostly
singlet, with a very  small admixture of $\nu_L^c$,
$L$ decays about half the time to a final
state containing a $W^\pm$ :  
  $BR( L \to W^+ e^-) \simeq 
BR( N_R \to H^+ e^-) = 
1/4$ \cite{Branco:2001pq}. This arises because the $1/m_W^2$
in the $W$ spin sum
\beq
S^{\mu \nu} = -g^{\mu \nu} + \frac{q^\mu q^\nu}{m_W^2}
\eeq
cancels the $g^2 v^2$ that appears upstairs
in the  squared matrix element:
\beq
\frac{g^2}{4} \left(\frac{\lambda v}{M_R} \right)^2
{\rm {\bf Tr}}
\{ \pslash_L \gamma_\mu \kslash_e \gamma_\nu \} S^{\mu \nu} ~~~.
\eeq
This confirms that  at high energy,
in the SSB theory,  rates involving 
the $W$ are similar to those for
the goldstone in the unbroken theory. 
  We therefore  assume
that NSI  operators induce $\bar{q} q \to
W^+ W^- e_\a^+ e^-_\b$  at the same
rate as     $\bar{q} q \to
H^+ H^- e_\a^+ e^-_\b$.

\subsection{The production cross-section}
\label{sigmaLHC}

In the Equivalence Theorem limit, where
the $W^\pm$ is approximated as a scalar
$H^\pm$, 
the matrix element for $ \bar{q} q \to 
e_\a^+ e_\b^- H^+ H^-$ is comparatively simple,
and  the partonic cross-section is
\beq
\hat{\sigma}(q \bar{q}  \to H^+ H^- e_\a^+ e_\b^-)
=
\frac{1}{15}
\frac{1}{2\hat{s}}
\left[
\frac{ (2 \pi)^4  \hat{s}^2}{\Lambda_8^8}
\right]
\left( \frac{2 \hat{s}^2}{3(4 \pi)^9}
\right)
\label{seceff}
\eeq
This result was obtained by
analytically integrating over massless
four body phase space up to the last
two integrals, which are performed by
Maple${}^{\mathrm{TM}}$ \cite{maple}. It corresponds to 1/15 of
a dimensional analysis estimate:
the last parenthese
is    massless four-body
phase space \cite{4pps},
and the square brackets  enclose
an estimate of
 the matrix element squared, 
together  with a $(2 \pi)^4$ which
accompagnied  the
four-momentum $\delta$-function.

\begin{figure}[ht]
\unitlength.5mm
\begin{center}
\includegraphics[scale=.4]{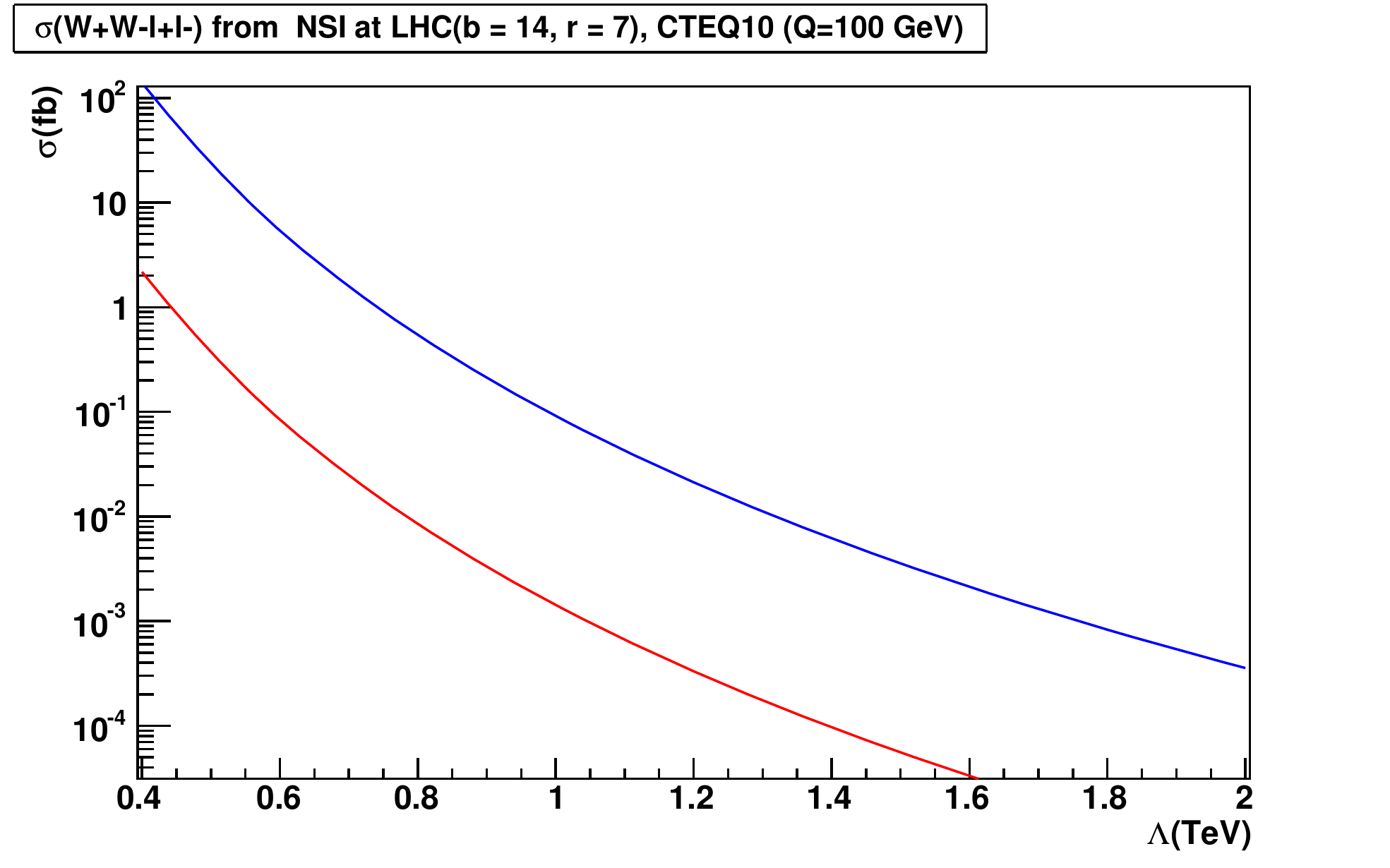}
\end{center}
\caption{ 
Estimated  cross-section $\sigma (pp \to W^+W^- e_\a^+
e^-_\b)$ in fb, at the LHC with 14 TeV (blue) and 7 TeV (red) centre of mass energy,
due to  dimension eight contact interactions with
coefficient $1/\Lambda^4$. 
\label{fig:sigd8} }
\end{figure}

Integrating  (\ref{seceff}) over  CTEQ10 parton distribution
functions\cite{CT10}, assuming 
the contact interaction 
couples equally to  $u$ and $d$ quarks,
 we obtain 
$\sigma(p  p \to
H^+H^-e_\a^+e^-_\b)$ at the 14 TeV
LHC plotted in figure \ref{fig:sigd8}. 
The  cross-section is small --- 
suppressed
by  four body  phase space---
so the LHC  could have difficulty detecting  NSI
that  appear as contact interactions.

We obtain  $\sigma (pp \to W^+W^-e^+_\a e^-_\b)$
``by hand'', via the Equivalence Theorem, because it is
straightforward and manageable.  We are unclear on
the coefficient and structure of the correspondingg
$\bar{q} q  W^+W^-e^+_\a e^-_\b$ interaction
which would arise in the spontaneously broken
electroweak theory,  and which could be
implemented in programmes such as Madgraph 
\cite{MadGraph5}. The Madgraph crosssections
we obtain, using the interactions
discussed in the following sections,
are consistent\footnote{This agreement
may be partly accidental. The ``longitudinal
$W$'' component of  the
$|$matrix element$|^2$ for the vertex of
Eqn. (\ref{effLHC}) is $(p_{W^+} \cdot p_{W^-})^2/m_W^4$
$\times$  the $|$matrix element$|^2$ for
$(\overline{q} \gamma^\mu q)(\overline{e_\a} \gamma_\mu e_\b)
H^+ H^-$. If Eqn. (\ref{eqt}) implies that these 
 $|$matrix element$|^2$s should be equal, then
the coefficient of  Eqn. (\ref{effLHC}) 
 has the curious value $m_W^2/(\Lambda_8^4 p_{W^+} \cdot p_{W^-})$.
Using that the average $\langle \hat{s} \rangle$
in the $ pp \to e_\a^+ e_\b^- H^+ H^-$ process is
 $\langle \hat{s} \rangle \sim 18$ TeV$^2$,
and that $\hat{s} = (p_{W^+} +  p_{W^-} + p_{e^+} +  p_{e^-})^2$,
we guess $ p_{W^+} \cdot p_{W^-} \simeq 1.5$ TeV$^2$. 
With this value, the Madgraph cross-section for the
interaction of  Eqn.  (\ref{effLHC})  agrees with
figure \ref{fig:sigd8}. \\
In the model of section \ref{secZ'}, we also
obtain a suppression of the 
$\bar{q} q  W^+W^-e^+_\a e^-_\b$ interaction
with respect to the coefficient of the gauge
invariant NSI operator. This $2m_W^2/M_R^2$
factor has a fixed mass in the denominator
(this is reassuring). However, $M_R$ is not
determined by $\Lambda_8$, and we do not expect to
need additional details of the model
to obtain $\sigma (q \bar{q}  \to e_\a^+ e_\b^- W^+W^-)$
in the Equivalence Theorem limit. Setting $M_R = \Lambda \sim$ TeV,
gives a coefficient numerically similar to
the  curious estimate above for the
coefficient of Eqn  (\ref{effLHC}), but this
could be an accident. Notice that we did not
compute the  $|$matrix element$|^2$ for Eqn.
 (\ref{effLHC}). }
 (to within  a factor
of a few) with figure \ref{fig:sigd8}.

\subsection{Detecting NSI contact interactions  at the LHC }
\label{eff}

To determine the reach of the LHC  in $\epsilon_{\a \b}$,
we need to discuss  backgrounds. 
The SM  cross-section for 
$ W^+ W^- \ell_\a^+ \ell_\a^-$ is about 
2 fb at $\sqrt{s} =  14$ TeV. 
  Besides irreducible backgrounds, 
one should also consider backgrounds faking our signal. 
For example, as discussed in  \cite{Delgado},
$W^\pm \, Z/\gamma^*$ 
 and  jets
has a  cross-section of 30 fb at the 14 TeV LHC,
and could look like 
$ W^+ W^- \ell_\a^+ \ell_\b^-$ when
the jets can be reconstructed to a $W^\pm$. 
More overwelming could be $t \bar{t}$+jets,
which could produce signals 
such as trileptons, where both $W$'s decay 
leptonically and one of the $b$-jets fakes an 
isolated lepton. To study the effect of a $b$-jet faking a 
lepton we took a 3M  $t \bar{t}$ sample generated with 
ALPGEN~\cite{ALPGEN} using MLM matching~\cite{MLM} 
and then showered with PYTHIAv6.4 \cite{pythia} at 14 TeV. 
The cross sections for $t \bar{t}$ + (0,1,2) 
jets are (440,778,730) pb. 
The efficiency of asking for 3 leptons 
in the sample is about 5$\times 10^{-3}$, 
which is consistent with the estimates in \cite{sullivan}. 
This effect would reduce the $t \bar{t}$ background 
to levels of 100's of fb, 
still too large for most of the NSI parameter space
(see figure \ref{fig:sigd8}).

Fortunately, asking for multi-leptons is not our only 
handle on reducing backgrounds. 
We anticipate that  an effective operator will 
produce highly boosted objects, 
allowing an  efficient cut on $p_T$,
and  also that the NSI events may be spherical.
To  study these expectations,
we simulate the  simple interaction
\beq
(\overline{q} \gamma^\r P_L q) (W^+_{\mu} \overline{\ell }^-_\a \gamma_\r W^{-,\mu} \ell^+_\b) 
\label{effLHC}
\eeq
 using FeynCalc~\cite{FeynCalc} and MadGraph v5~\cite{MadGraph5}. 
 $q$ can be
a $u$ or $d$ quark, and   both are assumed to couple with
 the same strength.  
 We impose parton level cuts of $|\eta|<$ 2.5 and $p_T>$ 20 GeV.
If we attribute to this interaction
the coefficient of the interaction (\ref{effLHC2}),
with $M_R = \Lambda_8$, MadGraph gives
the same total cross-section  as section
\ref{sigmaLHC}.

\begin{figure}[h]
\centering
\includegraphics[scale=.35]{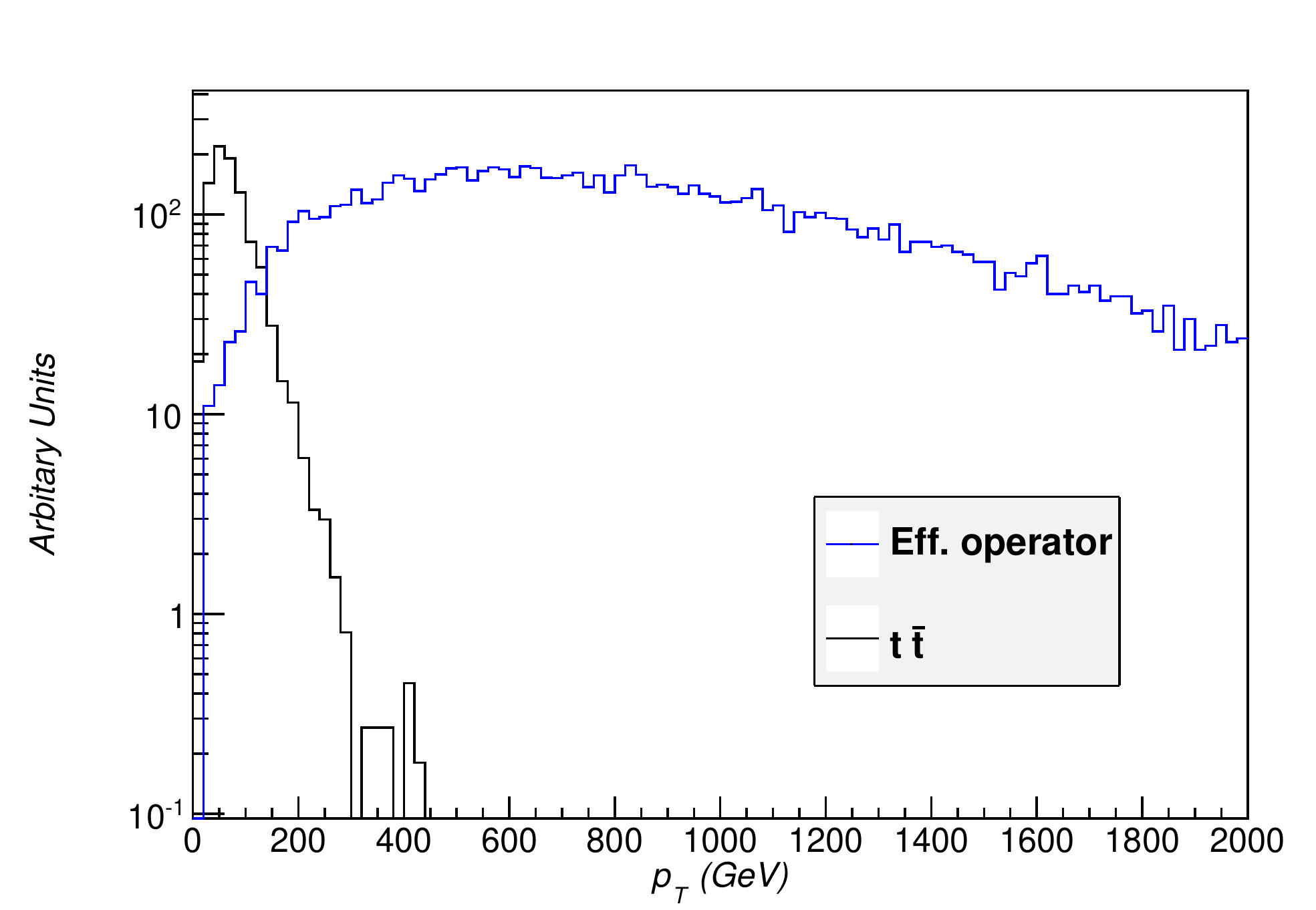}
\caption{  Distribution of $p_T$ of the leptons in NSI (blue) and $t \bar{t}$ (black).  The normalization of the distributions is chosen to show both signal and background in the same scale, and the energy of the LHC is 14 TeV. Note that the vertical axis is logarithmic.}
\label{fig:pT}
\end{figure}

In 
 Fig.~\ref{fig:pT} we plot the $p_T$ 
of objects in both the $t \bar{t}$ and signal samples. 
The signal and background distributions are well separated, 
and asking for a cut on $p_T$ of both leptons of 
order 400 GeV would reduce the $t \bar{t}$ backgrounds
to less 
than $10^{-5}$ fb and keeps 70\% of our signal.
We can interpret that the 
$\hat{s}^3$ dependence of eqn
(\ref{seceff})
 encourages energetic
final states. Indeed, integrating $\hat{s}$ under
the cross-section gives an average
$\sqrt{s}$ for NSI events of  order four TeV, or one TeV per particle, in agreement with the $p_T$ distribution obtained in the simulation.

The $p_T$  cut  is very effective 
for removing the  SM background. 
So one
 could relax 
the assumption on the number of electrons and muons, 
and search  for NSI contact interactions  by requiring 
 high $p_T$ multiparticle events.
In particular, this would give
sensitivity to NSI operators involving taus. 
However, if such events were seen, 
boosted taus would be hard to tag, 
although they  may show up as fat jets.

\begin{figure}[h]
\centering
\includegraphics[scale=.35]{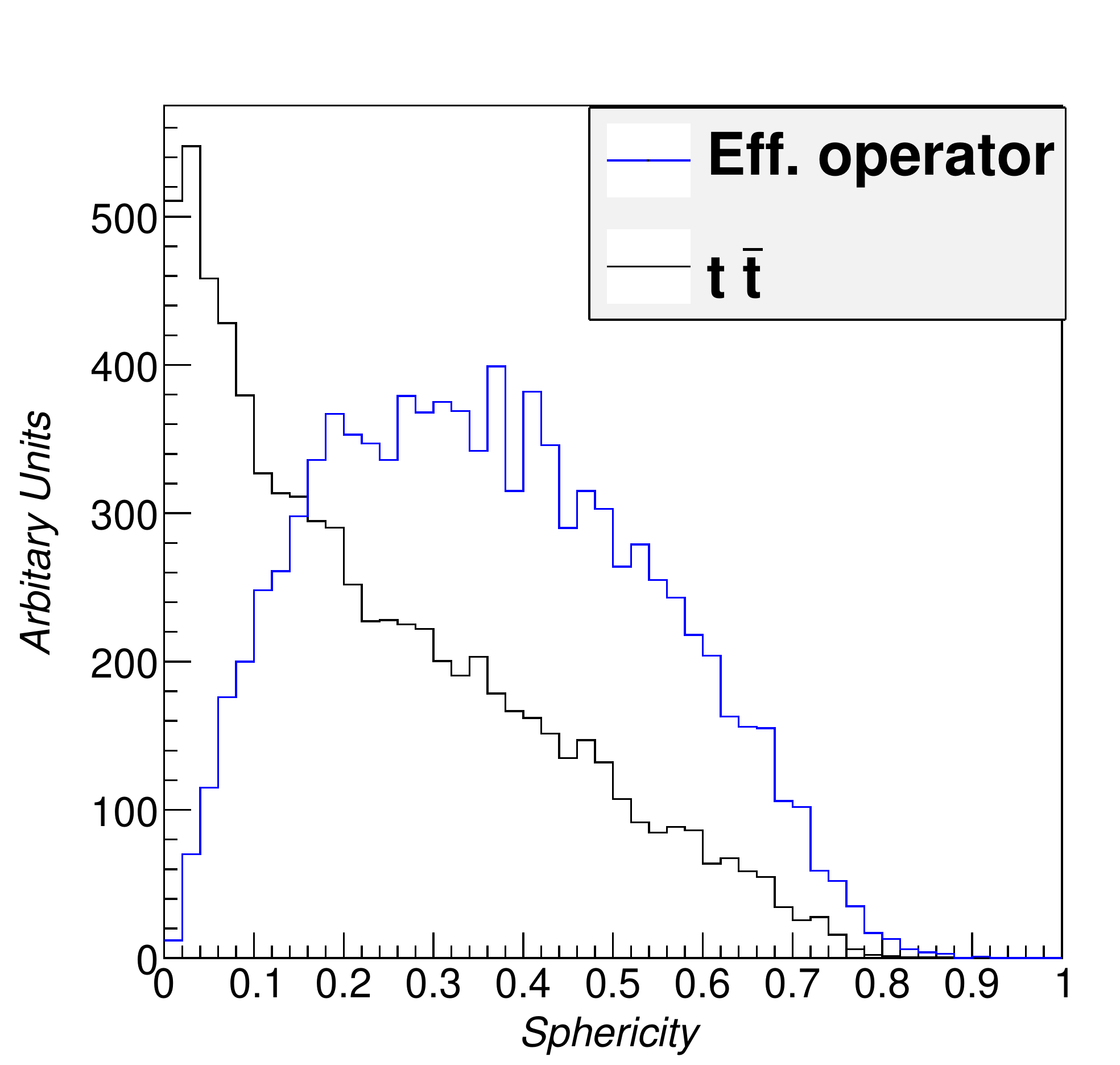}
\caption{ Sphericity in the signal (blue) and $t \bar{t}$ (black). The normalization of the distributions is chosen to show both signal and background in the same scale.}
\label{spher}
\end{figure}

In Fig.~\ref{spher} we show the sphericity of the signal versus the $t \bar{t}$ sample. As expected, the signal is more spherical than the background.
The signal objects are well separated in $\Delta$R space, so no issues of overlapping between the leptons and the $W$'s arise. The signal is not particularly forward: the effect of $|\eta_{\ell}|< 4.5$ to 2.5 is reducing the signal by less than 8\%. 

Summarizing, the cross-section is small,
 but the  signal is characterized by well-separated, 
highly boosted objects in a pretty spherical event. 
Using these characteristics, especially a cut on $p_T$, 
we showed that the largest background,
from $t \bar{t}$ can be reduced  below the 
signal. The LHC reach depends on the value of $\varepsilon$, 
but if we assume the NSI signal  is
background-free, and  ask for 100 events at a luminosity ${\cal L}$ 
(in fb$^{-1}$),  then the reach in $\varepsilon$ is
\bea
\varepsilon \sim 3 \times  10^{-2}/\sqrt{{\cal L}}
\label{LHCbds}
\eea
For example, for a luminosity of 100 fb$^{-1}$,  
the  14 TeV LHC  could be sensitive to 
NSI-induced contact interactions corresponding to 
\bea
\varepsilon \gsim  3 \times  10^{-3} ~~~.
\eea
This estimate can  be compared with  the 
$\varepsilon > 10^{-4}$ sensitivity of a
neutrino factory. 

So far we have discussed the LHC run at 14 TeV, but one can already obtain bounds using the current run at 7 TeV. The reduction in cross section from 14 to 7 TeV leads to a decrease in cross section of about a factor 60. Therefore, assuming the current running would end up with 30 fb$^{-1}$ of combined ATLAS+CMS data, the sensitivity with the 2011-12 data set would be $\epsilon \simeq 4 \times 10^{-2}$.

\subsection{A Z' model}
\label{secZ'}

In this section, we discuss how  an interaction
 similar to  Eq.~\ref{effLHC} could arise in a toy  model 
for NSI, and compare  this $Z'$ model
predictions  with the study done in the previous section. 

\begin{figure}[h]
\centering
\includegraphics[scale=.55]{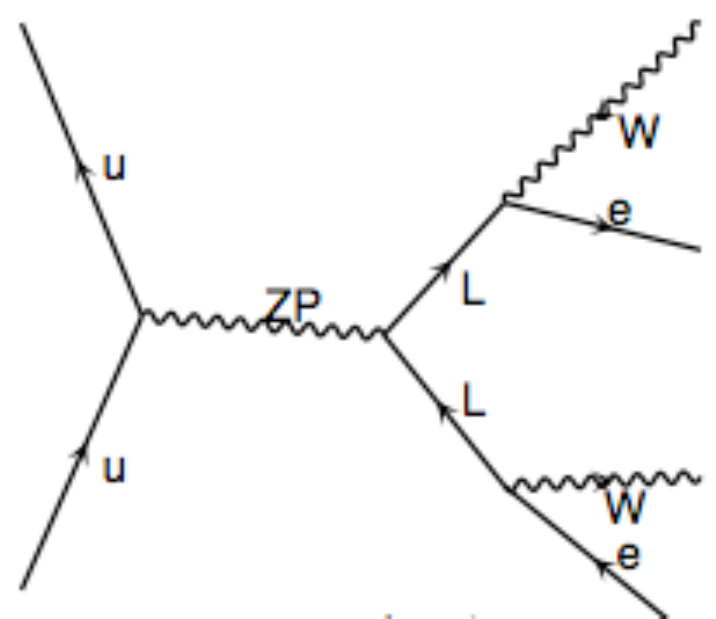}
\caption{ The high energy  
origin of $ pp \to e_\a^+ e^-_\b W^+ W^-$ in a Z' model with new leptons.}
\label{ZtoNN}
\end{figure}

Let us assume a new $Z'$, with coupling
$g_q' = 1$  
to first generation quarks,
so it is produced at hadronic colliders
via  $q \bar{q} \to Z'$. 
We suppose  a 
cancellation at the level of 
dimension-6 operators (as in Gavela et al. \cite{Gavela}), 
so neglect the $Z'$ contribution to
the dijet  cross section. 
Our $Z'$  also couples,  with
$g_N' = 1$,  to
new  singlet leptons $N_R$ , of mass
$M_R$, which have a
yukawa coupling  $\lambda$ to SM leptons
\footnote{It is unlikely that this toy model
is consistent with  low energy neutrino data.}.
After electroweak symmetry breaking,
the heavy mass eigenstate $L$ will
contain a small admixture of doublet
neutrino (see eqn (\ref{Leig})), and will 
have a coupling $\propto \lambda v/M_R$ to  $W$s. 
This allows the process illustrated in 
 Fig.~\ref{ZtoNN}. In the contact interaction
limit for  the $Z'$ and $L$ propagators, 
this diagram gives the interaction
\beq
\frac{g^2 v^2}{\Lambda^4 M_R^2}
(\overline{q} \gamma^\r P_L q) ( \overline{\ell }_\a  \Wslash
\gamma_\r \Wslash \ell_\b)  ~~~,
\label{effLHC2}
\eeq
which gives a somewhat more complicated matrix element
than the interaction simulated in the previous
section (see eqn  (\ref{effLHC})).

\begin{figure}[t]
\centering
\includegraphics[scale=.35]{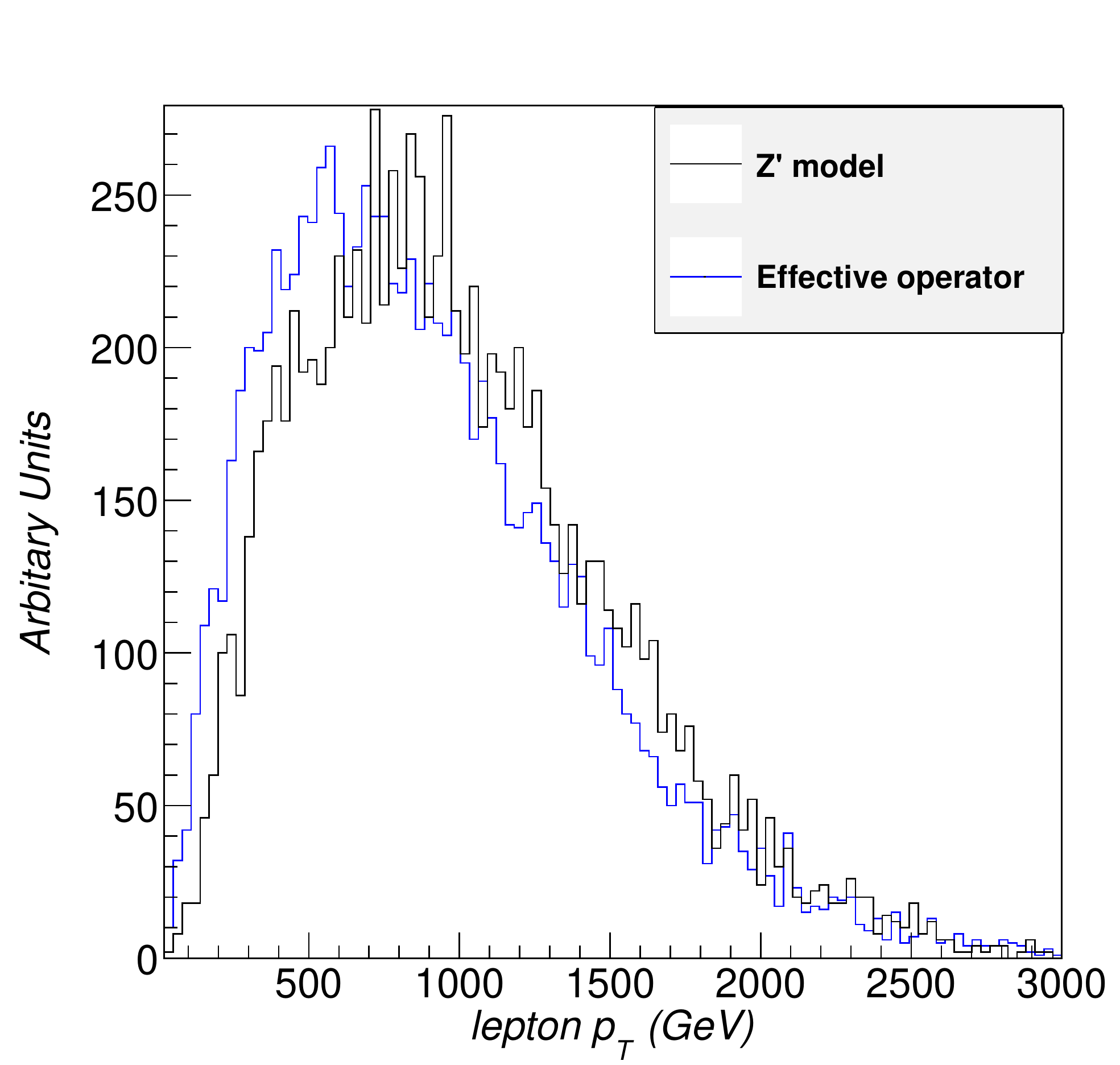}
\includegraphics[scale=.35]{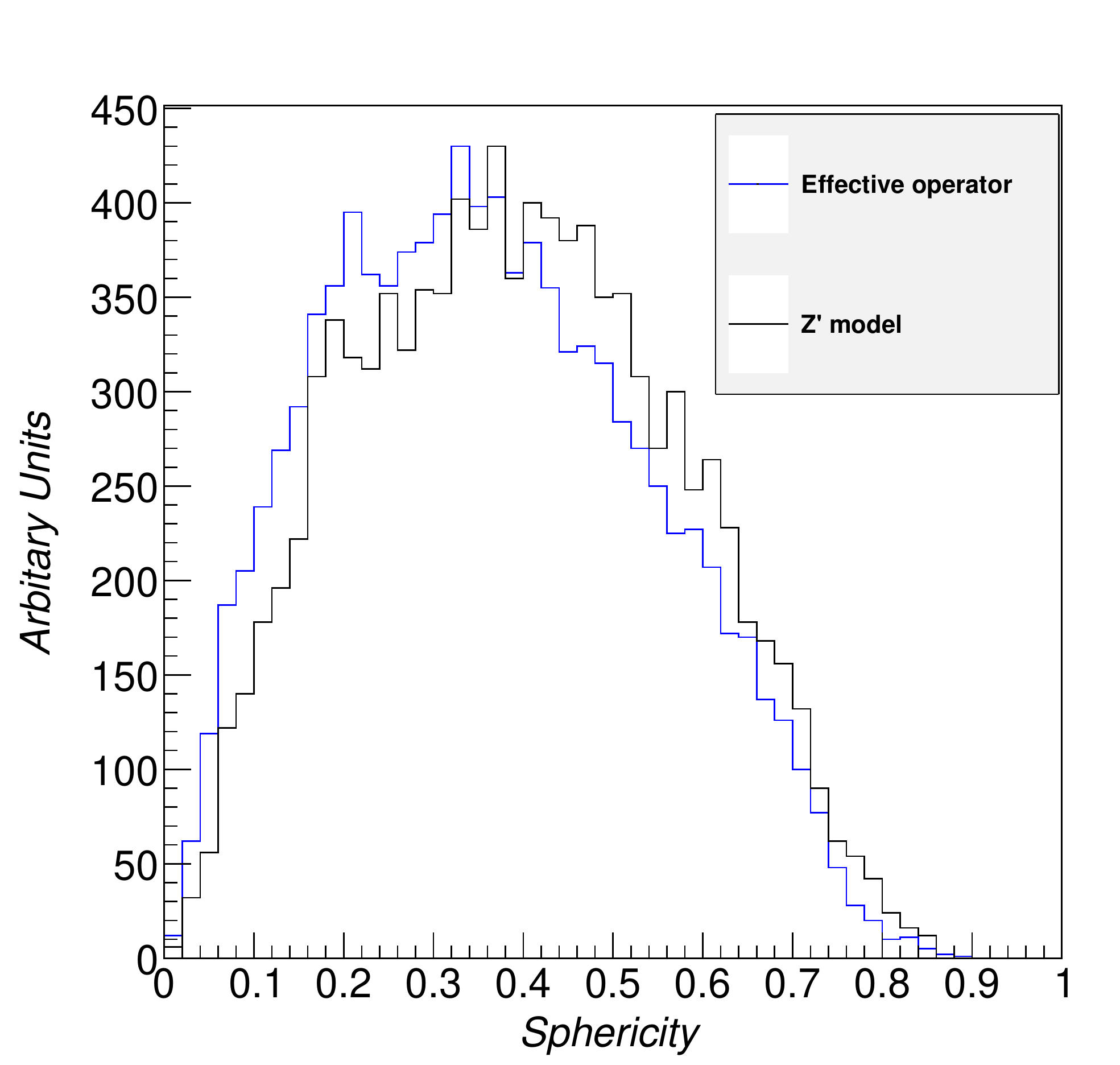}
\caption{ Comparison of the kinematic distributions for the Z' model and the effective operator.}
\label{effvsZp}
\end{figure}

We simulated the process
of  Fig.~\ref{ZtoNN}
 with MadGraph, 
requiring 
a large  mass for the Z' resonace, to ensure
that it is produced off-shell. The $L$
mass was allowed to vary in the 
$v \to $ TeV range. This allows to explore
an example where some of the new
particles responsable for NSI 
(the $N_R$), are kinematically
accessible to the LHC. 
The final state  kinematics for this process
are very similar 
to those of the effective operator analysis 
in the previous section. 
In Fig.~\ref{effvsZp} we show the $p_T$ 
distribution and sphericity for the effective 
operator and a model with a 3 TeV Z' and lighter $N_4$ with mass of 300 GeV.  

Although the kinematic distributions are very similar in the $Z'$ model and the contact interaction model, the cross sections are dramatically different. Indeed, after integration over PDFs, the cross sections are roughly related by the two-body versus four-body phase factors\cite{4pps}, 
\bea
\sigma_{Z'} \simeq 10^{5} \times \sigma_8
\eea
which would lead to a larger NSI signal. If we ask for 100 events at a luminosity ${\cal L}$ 
(in fb$^{-1}$),  then the reach in $\varepsilon$ is
\bea
\varepsilon \sim 10^{-4}/\sqrt{{\cal L}}
\label{LHCbds2}
\eea
which for a luminosity of 100 fb$^{-1}$ would lead to a sensitivity in the range of
$\varepsilon \gsim  10^{-5}$.

\section{Summary/Conclusion}

In this paper, we studied some
implications at colliders of  neutral current,
Non-Standard neutrino Interactions (NSI) of the
form of eqn (\ref{eqndisc}).  We assumed
these contact interactions to be the remnants, after
elextroweak symmetry breaking,
of gauge-invariant
dimension eight operators  such as eqn (\ref{eqn2}).
Notation and models that could give
such interactions are reviewed in section \ref{notn}.

In section \ref{LEP2}, we considered  the class
of models where 
\begin{itemize}
\item these operators are induced at
tree level  by New Physics 
\item this New Physics  does
not induce ``dangerous'' dimension six operators
involving charged leptons, 
because the various contributions to
their coefficients cancel.
\end{itemize}
We argued that this cancellation only obtains at zero 
momentum transfer, so  the New Physics could
induce  contact  interactions involving  charged
leptons rather than neutrinos, with coefficients $\sim s/\Lambda_8^4$
(where $s$ in the centre of mass energy).
These contact interactions  must satisfy
LEP2 (and Tevatron/LHC) bounds, which
can be translated via eqn (\ref{quesuisje1})
into approximate bounds 
$\varepsilon \lsim 10^{-2}, 10^{-3}$ on
the coefficient of NSI.  The bounds on
NSI  operators  involving 
$(\bar{e} e)(\overline{\ell_a} \ell_a)$,
 $(\bar{e} e)(\overline{\ell_a} \ell_b)$
for $\a \neq \b$, and  $(\bar{e} e)(\overline{q} q)$,
are given respectively in tables
\ref{tab:LEP2bds}, \ref{tabFV}, and
\ref{tab:qqll}. It is possible
to constrain flavour-changing four-lepton
interactions because OPAL published bounds\cite{OPALFV}
on contact interactions giving $e^\pm \mu^\mp$,
 $\tau^\pm \mu^\mp$, and  $e^\pm \tau^\mp$ in
the final state.

In section \ref{LHC}, we focussed  on NSI that
would behave like a contact interaction  at
LHC energies.  We  used the Equivalence Theorem 
to relate NSI operators involving  Higgses and neutrinos, 
with operators with two charged gauge bosons and
two charged leptons. The cross-section we obtain
 for  
$pp \to W^+ W^- e^+_\a e^-_\b$ 
is small (see figure \ref{fig:sigd8}), due to four-body final state
phase space suppression, but gives
spherical 
events with highly-boosted objects. On the other hand,
section \ref{secZ'} discusses a $Z'$ model  where some of the
new particles can be produced on-shell,
which enhances the
the cross section  by orders of magnitude. 
Interestingly,  the final state kinematics  remain 
similar. 

We showed in
section \ref{eff}  that 
asking for multileptons, and in particular, imposing  a 
hard cut on $p_T$ (see figure \ref{fig:pT}), 
can reduce  Standard Model backgrounds
below the signal.  This is even the case
for NSI involving $\tau$s. Our results suggest
that  the LHC with 100 fb$^{-1}$ of data,
could be sensitive to contact interactions
induced by NSI with $\varepsilon \gsim 3 \times 10^{-3}$ ($\varepsilon \gsim 10^{-5}$) for the model independent (dependent) case,
(see eqns (\ref{LHCbds}) and (\ref{LHCbds2})).

\vspace{0.5cm}
\begin{acknowledgments}
VS would like to thank Claude Duhr and C\'eline Degrande for helping on the FeynCalc implementation of the effective operator.
VS' research is partly funded by NSERC funds. SD  thanks the York University
Physics Dept for hospitality when this
work was begun, and  NSERC for  visitor funding.
\end{acknowledgments}

\end{document}